\documentclass[[a4paper,12pt]{article}
\usepackage{fullpage}
\newcommand{\keywords}[1]{\par\medskip\noindent\textbf{Keywords:} #1}
\newcommand{\email}[1]{\\texttt{#1}}
\newcommand{\codefigsize}{\normalsize}
\usepackage[english]{babel}
\usepackage[utf8]{inputenc}
\usepackage{xcolor}
\usepackage{amsmath,amssymb}
\usepackage{listings}
\usepackage{xspace}
\usepackage{doi}
\usepackage{todonotes}
\definecolor{gray}{rgb}{0.6,0.6,0.6}
\definecolor{mauve}{rgb}{0.58,0,0.82}
\definecolor{auburn}{rgb}{0.43, 0.21, 0.1}
\definecolor{babyblue}{rgb}{0.54, 0.81, 0.94}
\definecolor{amaranth}{rgb}{0.9, 0.17, 0.31}
\definecolor{bleudefrance}{rgb}{0.19, 0.55, 0.93}
\definecolor{atomictangerine}{rgb}{1.0, 0.6, 0.4}
\definecolor{beaublue}{rgb}{0.74, 0.83, 0.9}
\definecolor{dkviolet}{rgb}{0.9, 0.17, 0.31}
\definecolor{dkgreen}{rgb}{0.0, 0.42, 0.24}
\definecolor{ltblue}{rgb}{0.0, 0.75, 1.0}
\definecolor{dkblue}{rgb}{0.2, 0.2, 0.6}
\definecolor{dkred}{rgb}{0.8, 0.0, 0.0}
\definecolor{byzantine}{rgb}{0.74, 0.2, 0.64}
\lstdefinestyle{bsmlstyle}{
language=[Objective]caml,
mathescape=true,
texcl=false,
morekeywords=[4]{bsp_p,bsp_g,bsp_l,bsp_s,mkpar,apply,proj,put,par,Bsml},
morekeywords=[2]{int,bool,float,unit,char,string,list,option,array},
morecomment=[s]{(*}{*)},
showstringspaces=false,
morestring=[b]",
morestring=[d],
tabsize=2,
extendedchars=false,
sensitive=true,
breaklines=false,
basicstyle=\ttfamily,
captionpos=b,
basewidth={2em, 0.5em},
columns=[l]flexible,
numberstyle=\sffamily\tiny\color{gray},
identifierstyle={\ttfamily\color{black}},
keywordstyle=[1]{\ttfamily\bfseries\color{dkviolet}},
keywordstyle=[2]{\ttfamily\bfseries\color{dkgreen}},
keywordstyle=[3]{\ttfamily\bfseries\color{ltblue}},
keywordstyle=[4]{\ttfamily\bfseries\color{dkblue}},
keywordstyle=[5]{\ttfamily\color{dkred}},
keywordstyle=[6]{\ttfamily\color{dkpink}},
stringstyle=\ttfamily,
commentstyle={\ttfamily\itshape\color{gray}},
numberstyle=\tiny\color{gray},
literate=
        {'a}{{\color{dkgreen}{$\alpha$}}}1
{'b}{{\color{dkgreen}{$\beta$}}}1
{'c}{{\color{dkgreen}{$\gamma$}}}1
{'d}{{\color{dkgreen}{$\delta$}}}1
{'e}{{\color{dkgreen}{$\epsilon$}}}1
{->}{{\color{dkviolet}{$\rightarrow$}}}2
{=}{{\color{dkviolet}{$=$}}}1
}
\lstdefinelanguage{Coq}{
mathescape=true,
texcl=false,
morekeywords=[1]{Section, Module, End, Require, Import, Export,
                Variable, Variables, Parameter, Parameters, Axiom, Hypothesis,
                Hypotheses, Notation, Local, Tactic, Reserved, Scope, Open, Close,
                Bind, Delimit, Definition, Let, Ltac, Fixpoint, CoFixpoint, Add,
                Morphism, Relation, Implicit, Arguments, Unset, Contextual,
                Strict, Prenex, Implicits, Inductive, CoInductive, Record,
                Structure, Canonical, Coercion, Context, Class, Global, Instance,
                Program, Infix, Theorem, Lemma, Corollary, Proposition, Fact,
                Remark, Example, Proof, Goal, Save, Qed, Defined, Hint, Resolve,
                Rewrite, View, Search, Show, Print, Printing, All, Eval, Check,
                Projections, inside, outside, Def},
morekeywords=[2]{forall, exists, exists2, fun, fix, cofix, struct,
                match, with, end, as, in, return, let, if, is, then, else, for, of,
                nosimpl, when},
morekeywords=[3]{Type, Prop, Set, true, false, option},
morekeywords=[4]{pose, set, move, case, elim, apply, clear, hnf,
                intro, intros, generalize, rename, pattern, after, destruct,
                induction, using, refine, inversion, injection, rewrite, congr,
                unlock, compute, ring, field, fourier, replace, fold, unfold,
                change, cutrewrite, simpl, have, suff, wlog, suffices, without,
                loss, nat_norm, assert, cut, trivial, revert, bool_congr, nat_congr,
                symmetry, transitivity, auto, split, left, right, autorewrite},
morekeywords=[5]{by, done, exact, reflexivity, tauto, romega, omega,
                assumption, solve, contradiction, discriminate},
morekeywords=[6]{do, last, first, try, idtac, repeat},
morecomment=[s]{(*}{*)},
showstringspaces=false,
morestring=[b]",
morestring=[d],
tabsize=2,
extendedchars=true,
sensitive=true,
breaklines=false,
captionpos=b,
basewidth={2em, 0.5em},
columns=[l]flexible,
basicstyle={\ttfamily\color{black}},
numberstyle=\sffamily\tiny\color{gray},
identifierstyle={\ttfamily\color{black}},
keywordstyle=[1]{\ttfamily\bfseries\color{dkviolet}},
keywordstyle=[2]{\ttfamily\bfseries\color{dkgreen}},
keywordstyle=[3]{\ttfamily\bfseries\color{ltblue}},
keywordstyle=[4]{\ttfamily\color{dkblue}},
keywordstyle=[5]{\ttfamily\color{dkred}},
stringstyle=\ttfamily,
commentstyle={\ttfamily\itshape\color{dkgreen}},
literate=
        {forall}{{\color{dkgreen}{$\forall\;$}}}1
{exists}{{$\exists\;$}}1
{<-}{{$\leftarrow\;\;$}}1
{=>}{{$\Rightarrow\;\;$}}1
{->}{{$\rightarrow\;\;$}}1
{<-->}{{$\longleftrightarrow\;\;$}}1
{<->}{{$\leftrightarrow\;\;$}}1
{<==}{{$\leq\;\;$}}1
{\#}{{$\#$}}1
{∘}{{$\circ\;$}}1
{\\o}{{$\circ\;$}}1
{\/\\}{{$\wedge\;$}}1
{\\\/}{{$\vee\;$}}1
{~}{{\ }}1
{\@\@}{{$@$}}1
{\\mapsto}{{$\mapsto\;$}}1
{\\hline}{{\rule{\linewidth}{0.5pt}}}1
}[keywords,comments,strings]
\lstdefinestyle{coqstyle}{language=Coq}
\lstdefinelanguage{why}{
mathescape=true,
texcl=false,
morekeywords=[1]{module,end,let,rec,match,with,fun,use,val,axiom,theory,lemma,predicate,type,ghost,function,if,then,else,import,export,raise},
morekeywords=[2]{variant,requires,ensures,returns,assert,absurd,invariant,raises},
morekeywords=[3]{int,list,bool,option,Int,List,Length,NthNoOpt,Option},
morekeywords=[4]{true,false},
morekeywords=[5]{},
morekeywords=[6]{},
morecomment=[s]{(*}{*)},
showstringspaces=false,
morestring=[b]",
morestring=[d],
tabsize=2,
extendedchars=true,
sensitive=true,
breaklines=false,
captionpos=b,
basewidth={2em, 0.5em},
columns=[l]flexible,
basicstyle={\ttfamily\color{black}},
identifierstyle={\ttfamily\color{black}},
keywordstyle=[1]{\ttfamily\bfseries\color{dkviolet}},
keywordstyle=[2]{\ttfamily\bfseries\color{blue}},
keywordstyle=[3]{\ttfamily\bfseries\color{dkgreen}},
keywordstyle=[4]{\ttfamily\color{dkblue}},
keywordstyle=[5]{\ttfamily\color{dkred}},
stringstyle=\ttfamily,
commentstyle={\ttfamily\itshape\color{olive}},
numberstyle=\sffamily\tiny\color{gray},
literate=
        {forall}{{\bfseries\color{blue}{$\forall\;$}}}1
{exists}{{\bfseries\color{blue}$\exists$}}1
{<-}{{\bfseries\color{blue}$\leftarrow$}}1
{<>}{{$\not=$}}1
{>=}{{$\geq$}}2
{<=}{{$\leq$}}2
{->}{{\bfseries\color{blue}$\rightarrow$}}2
{<->}{{\bfseries\color{blue}$\leftrightarrow\$}}3
{\/\\}{{\bfseries\color{blue}$\wedge$}}2
{\\\/}{{\bfseries\color{blue}$\vee$}}2
{\@\@}{{\bfseries\color{blue}$@$}}1
{'a}{{\color{dkgreen}{$\alpha$}}}1
{'b}{{\color{dkgreen}{$\beta$}}}1
{'c}{{\color{dkgreen}{$\gamma$}}}1
{'d}{{\color{dkgreen}{$\delta$}}}1
{'e}{{\color{dkgreen}{$\epsilon$}}}1
}[keywords,comments,strings]
\lstdefinestyle{whystyle}{language=why}
\lstMakeShortInline[style=bsmlstyle]"
\lstMakeShortInline[style=coqstyle]!
\newcommand{\w}[1]{\lstinline[style=whystyle]{#1}}

\newcommand{\why}{\textsc{Why3}\xspace}

\newcommand{\pvec}[2]{\langle #1,~\dots,~#2\rangle}

\makeatletter
\AtBeginDocument{%
  \def\doi#1{\url{https://doi.org/#1}}}
\makeatother
\begin{document}
\title{Verified Scalable Parallel Computing with \why}
\author{Olivia Proust \qquad Frédéric Loulergue\\
  Laboratoire d'Informatique Fondamentale d'Orléans\\
  Univ Orléans, INSA CVL, LIFO EA 4022, Orléans, France\\
  \email{olivia.proust2@gmail.com,~frederic.loulergue@univ-orleans.fr}
}
\maketitle
\begin{abstract}
  BSML is a pure functional library for the multi-paradigm language OCaml. BSML embodies the principles of the Bulk Synchronous Parallel (BSP) model, a model of scalable parallel computing. We propose a formalization of BSML primitives with WhyML, the specification language of Why3 and specify and prove the correctness of most of the BSML standard library. Finally, we develop and verify the correctness of a small BSML application.

\end{abstract}
\section{Introduction}
\label{sec:introduction}
High-level approaches to big data analytics such as Hadoop MapReduce~\cite{WHI2010:OR} or Apache Spark~\cite{ADD2015:PVLDB} are often inspired by bulk synchronous parallelism (BSP)~\cite{VAL1990:CACM} a model of scalable parallel computing. In this context, scalable means that the number of processors of the parallel machines running BSP programs could range from a few to several dozens of thousand cores or more. Bulk Synchronous Parallel ML (BSML)~\cite{LGB2005:ICCS} is a pure functional library for the multi-paradigm language OCaml\footnote{\url{https://ocaml.org}}. BSML embodies the principles of the BSP model, at a higher level than libraries such as the BSPlib library~\cite{HILL1998:BSPLIB} and can easily express patterns~\cite{GG2009:APDCM,LOU2017:PDCAT} (or algorithmic skeletons~\cite{COL1989:BOOK}) of frameworks such as MapReduce or Spark.

\why~\cite{BFM2014:STTT,why3_1_6} is a framework for the deductive verification of programs. It provides a specification and programming language named WhyML which can be used directly or as an intermediate language for other tools to verify C~\cite{KKP2015:FAC}, Java\cite{FM2007:CAV}, Ada or Rust~\cite{DJM2022:IFCEM} programs. The framework itself also provides mini-C and mini-Python front-ends. \why generates verification conditions to be verified by external provers. One of the strength of \why is that it targets a large variety of provers including Alt-Ergo~\cite{CCI2018:SMT}, Z3~\cite{MB2008:TACAS} and CVC5. Correct-by-construction OCaml code can be extracted from WhyML.

Our contributions are the formalization of BSML and its standard library in WhyML and its use in the specification and verification of a scalable parallel function for the maximum prefix sum problem, using map and reduce skeletons.

The remaining of the paper is organized as follows. In Section~\ref{sec:why}, we give an overview of \why and WhyML, including its limitations when dealing with higher-order functions. We introduce functional bulk synchronous parallel programming with BSML in Section~\ref{sec:fbsp}. Section~\ref{sec:bsmlwhy} is devoted to the formalization of the primitives of BSML and its application to the specification and verification of the BSML standard library. We consider the specification, development and verification of a small application: a parallel function that solves the maximum prefix sum problem in Section~\ref{sec:application}. We discuss related work in Section~\ref{sec:related_work} and conclude in Section~\ref{sec:conclusion}.

The set of \why modules is called WhyBSML and is available at \doi{10.5281/zenodo.8166092}.

\section{An Overview of \why}
\label{sec:why}
\begin{figure}[!h]
        \centering
        \begin{minipage}{.85\textwidth}
\begin{lstlisting}[style=whystyle,basicstyle=\codefigsize\ttfamily,numbers=left]
module Max
    use int.Int
        
    let max (x : int) (y : int) : int 
      ensures { result = x \/ result = y }
      ensures { result >= x /\ result >= y }
    = if x < y then y else x 
end

module MaxList
  use int.Int
  use list.List 
  use list.Length
  use list.NthNoOpt
  use Max

  function ([]) (l : list 'a) (i : int) : 'a = nth i l

  let rec maximum (l : list int) : int 
    requires { length l > 0 } 
    ensures { forall i:int. 0 <= i < length l -> result >= l[i] }
    ensures { exists i:int. 0 <= i < length l /\ l[i] = result }
    variant { l }
  = match l with 
    | Nil -> absurd
    | Cons h Nil -> h 
    | Cons h t -> 
      let _ = assert { forall i:int. 0 <= i < length t -> 
                                l[i+1] = t[i] } in
      max h (maximum t)
    end
end
\end{lstlisting}
        \end{minipage}
        \caption{A WhyML Example}
        \label{fig:why:maximum}‡
\end{figure}

\subsection{Specifying and verifying functional programs with \why}

\why is often used in the verification of imperative programs. As BSML is purely functional and BSML applications mostly used the functional features of OCaml, we focus here on the verification of functional programs. This focus is also a necessity as we will explain in the next subsection.

In addition to its core features, \why provides a standard library with data structures such as lists and arrays, as well as basic arithmetic logic with integers and reals. We illustrate this short introduction with the example of Figure~\ref{fig:why:maximum}. Note that this figure presents a pretty-printed version of the actual code, for example $\mathtt{/\backslash}$ is rendered as $\wedge$, \texttt{->} as $\rightarrow$, \texttt{'a} as $\alpha$, \textit{etc.}

WhyML developments are organized in \emph{modules}. The example defines two modules: \lstinline{Max} (lines 1--9) and \lstinline{MaxList} (lines 11-33). Defined modules can be used in other modules with the \lstinline{use} keyword. We use some modules of \why standard library: \lstinline{int.Int} about integer arithmetic (lines 2 and 12) and \lstinline{list.List}, \lstinline{list.Length}, \lstinline{list.NthNoOpt} for basic definitions and facts about lists (lines 13--15).

The module \lstinline{Max} is devoted to the specification and definition of a function \lstinline{max} which returns the largest of two integers. This function does not have any pre-condition but its post-conditions are introduced by the keyword \lstinline{ensures}.

Assuming the file \texttt{maximum.mlw} contains only the module \lstinline{Max}, verifying that \lstinline{max} satisfies its preconditions using the prover Alt-Ergo can be done with the following command:
{\codefigsize
\begin{verbatim}
why3 prove --prover alt-ergo maximum.mlw    
\end{verbatim}}
\noindent and the tool answers \lstinline{max} indeed satisfies its contract:
{\codefigsize
\begin{verbatim}
File maximum.mlw:
Goal max'vc.
Prover result is: Valid (0.00s, 8 steps).    
\end{verbatim}}

In our study, most of the functions to verify are recursive and often manipulate lists. Lines 19--30 are an example of a recursive function that takes a list of integers and returns the highest value the list contains.

To write the contract of function \lstinline{maximum}, we use the notation \lstinline{l[i]} to access the $i^\text{th}$ element of list \lstinline{l}. This notation is defined as a binary function in line~17 and is actually an alias for the \lstinline{nth} function of the standard library. Note that this definition is introduced by the keyword \lstinline{function} instead of the keyword \lstinline{let} (as in line 4). The purpose of \lstinline{([])} is to be used only in specifications while \lstinline{max} is code that is meant to be executed. Pure functions may be used in both roles if they are defined using both keywords. In this example, \lstinline{max} cannot be used in assertions while the bracket notation cannot be used in programs.

For \lstinline|maximum|, we have a larger contract with new clause types. We  add a pre-condition (following the keyword \lstinline{requires}) to this contract, due to the fact that our function is not defined on empty lists. To ensure termination, we define a \lstinline{variant}, which must be decreasing with each recursive call. The recursive call in line~30 is indeed called on the tail of the input list, thus this called is made on strictly smaller argument than \lstinline{l}.

We need quantifiers to express our post-conditions. The maximum value must be contained in the list (line 22 using \lstinline{exists}), and must be greater than or equal to all the values in the list (line 21 using \lstinline{forall}).

The definition of the function follows in lines 24--30. It proceeds by pattern matching on the input list. The case of the empty list (constructor \lstinline{Nil}) is \lstinline{absurd} as the pre-condition specifies the input list should not be empty (expressed as a fact on its length in line 20). If the list is a singleton (case \lstinline{(Cons h Nil)}), the result is of course the only element of the list. Otherwise --- and let us ignore lines 28--29 for the moment --- the result is the maximum of the head and the recursive call on the tail (line 30). Without lines 28--29, the execution of the tool now answers:
{\codefigsize
\begin{verbatim}
File maximum.mlw:
Goal max'vc.
Prover result is: Valid (0.00s, 8 steps).
File maximum.mlw:
Goal maximum'vc.
Prover result is: Timeout (5.00s).
\end{verbatim}}
Using Z3 or CVC5, or increasing the timeout, or changing the proof strategy does not change the outcome. It is possible to apply transformations to the goals. Using the \why IDE, just splitting the verification condition for \lstinline{maximum} gives five verification conditions: one for verifying the empty case is indeed absurd, one to check that the recursive call is indeed decreasing, one to check the pre-condition of the recursive call and one for each post-conditions. All these sub-goals are valid but the one corresponding to the post-condition in line 22 which remains unknown. To help the provers, we added lines 28--29 which relate elements of \lstinline{l} with elements of its tail via \lstinline{nth}. This assertion is easily verified and then eases the verification of the post-condition. The answer of the tool changes to:
{\codefigsize
\begin{verbatim}
Prover result is: Valid (0.09s, 749 steps).
\end{verbatim}}

\subsection{Limitations with higher-order functions}

\begin{figure}[!h]
        \centering
        \begin{minipage}{.85\textwidth}
\begin{lstlisting}[style=whystyle,basicstyle=\codefigsize\ttfamily,numbers=left]
module Concrete
  use export option.Option
      
  let remove_option (opt : option 'a) : 'a 
    requires { opt <> None } ensures { (Some result) = opt } 
  = match opt with
    | Some x -> x
    | None -> absurd
    end
end
module Failure
  use Concrete
  let apply (f:'a->'b)(a:'a) : 'b = f a

  let test_KO (c: option 'a) : 'a (* CANNOT BE VERIFIED *)
    requires { c <> None } ensures { Some(result) = c } 
  = apply remove_option c
end
module Abstract 
  use export option.Option
    
  val function remove_option(opt: option 'a) : 'a   
  axiom remove_option: forall x: 'a. remove_option(Some x) = x
end
module Success      
  use Abstract 
  let apply (f:'a->'b)(a:'a) : 'b = f a

  let test_OK (c: option 'a) : 'a
    requires { c <> None } ensures { Some(result) = c } 
   = apply remove_option c     
end
\end{lstlisting}
        \end{minipage}
        \caption{Limitations with Higher-Order Functions}
        \label{fig:why:higher_order}
\end{figure}

To show the limitations of \why in handling higher-order functions, let us consider the example of Figure~\ref{fig:why:higher_order}. Intuitively,  \lstinline{option 'a} extends the type \lstinline{'a} with a value \lstinline{None} and all the other values are encapsulated in the constructor \lstinline{Some}.

In lines 1--10, we define a module \lstinline{Concrete} containing the definition of a function \lstinline{remove_option} that extracts the value encapsulated in an optional value assuming this value is not \lstinline{None}. In the module \lstinline{Failure}, we apply this function but through a higher-order function \lstinline{apply} that just applies a function to a value. The tool fails to verify the function \lstinline{test_KO} which intuitively does exactly the same as \lstinline{remove_option}.
Note that if \lstinline{remove_option} was performing side effects or was partial because it may raise exceptions, \why would reject the program with an error. Here the problem is less visible. Indeed, the arguments of a higher-order function must be purely functional and \emph{total} functions.
In our case \lstinline{remove_option} is not total as its pre-condition excludes \lstinline{None}. The manifestation of the problem can be seen in a sub-verification condition generated by \why:
\lstinline{forall opt:option 'a. opt <> None}, which is impossible to prove.

Still, as most BSML primitives are higher-order functions, and we need to use functions such as \lstinline{remove_option}, a work-around was needed. Our solution is shown in module \lstinline{Abstract} (lines 19--24). Instead of writing a concrete implementation of \lstinline{remove_option}, we \emph{declare} a function \lstinline{remove_option} without defining it, and we only give its semantics (with an \lstinline{axiom}) when the pre-condition is met. It looks like a total function but if its application does not satisfy the precondition then it is impossible to reason about the result of the application. If the overall verification of a client code works despite an incorrect application of \lstinline{remove_option}, it means the result of the incorrect application was not used.
In module \lstinline{Success}, the same client code as module \lstinline{Failure} uses module \lstinline{Abstract} instead of module \lstinline{Concrete} and the verification succeeds.

\section{Functional Bulk Synchronous Parallelism}
\label{sec:fbsp}
The OCaml language is a versatile programming language that combines functional, imperative and object-oriented paradigms. BSML~\cite{LGB2005:ICCS} (Bulk Synchronous Parallel ML) is an OCaml-based library that embodies the principles of the BSP~\cite{VAL1990:CACM} (Bulk Synchronous Parallel) model. It provides a range of constants and functions to facilitate BSP programming. The BSP machine, viewed as a homogeneous distributed memory system with a point-to-point communication network and a global synchronization unit, serves as the underlying architecture for BSML. BSP programs, composed of consecutive super-steps, run on this kind of machine. The execution of each super-step follows a distinct pattern, starting with the computation phase where each processor-memory pair performs local computations using data available locally. This phase is followed by the communication phase, during which processors can request and exchange data with other processors. Finally, the synchronization phase concludes the super-step, synchronizing all processors globally.

With its collection of four expressive functions and constants like "bsp_p" representing the number of processors in the BSP machine, BSML empowers developers to create BSP algorithms. While OCaml supports imperative programming and BSML can exploit it~\cite{LOU2017:SCPE}, in this paper we only consider the pure functional aspects of OCaml and BSML. This deliberate focus differentiates it from the imperative counterparts provided by libraries such BSPlib for C~\cite{HILL1998:BSPLIB}. The types and informal semantics of BSML primitives are listed in Figure~\ref{fig:bsml:primitives}

\begin{figure}[!ht]
    \[
        \begin{array}{l}
            \mathtt{bsp\_p} : \mathtt{int}                                                                                      \\
            \mathtt{bsp\_p} = p
            \\[1.5mm]
            \mathtt{mkpar} : (\mathtt{int}\rightarrow\alpha)%
            \rightarrow \alpha\,\mathtt{par}                                                                                    \\
            \mathtt{mkpar~}f = \pvec{f\,0}{f\,(p-1)}                                                                            \\[1.5mm]
            \mathtt{proj} : \alpha\,\mathtt{par} \rightarrow %
            (\mathtt{int}\rightarrow\alpha)                                                                                     \\
            \mathtt{proj~}\pvec{v_0}{v_{p-1}} = %
            \mathtt{function}\,0\rightarrow v_0~|~\ldots~|~p-1 \rightarrow v_{p-1}                                              \\[1.5mm]
            \mathtt{apply} : (\alpha\rightarrow\beta)\mathtt{par}%
            \rightarrow %
            \alpha\,\mathtt{par} %
            \rightarrow %
            \beta\,\mathtt{par}                                                                                                 \\
            \mathtt{apply~}\pvec{f_0}{f_{p-1}}\ \pvec{v_0}{v_{p-1}} = %
            \pvec{f_0\ v_0}{f_{p-1}\ v_{p-1}}                                                                                   \\[1.5mm]
            \mathtt{put} : (\mathtt{int}\rightarrow\alpha)\mathtt{par}%
            \rightarrow %
            (\mathtt{int}\rightarrow\alpha)\mathtt{par}                                                                         \\
            \mathtt{put~} \pvec{\mathit{tosend}_0}{\mathit{tosend}_{p-1}} = \pvec{\mathit{received}_0}{\mathit{received}_{p-1}} \\
            \hfill \text{where for all } \mathit{src},\ \mathit{dst}                                                            \\
            \hfill 0\leq \mathit{src},\mathit{dst} < p \Rightarrow \mathit{received}_{\mathit{dst}}\ \mathit{src} = \mathit{tosend}_{\mathit{src}}\ \mathit{dst}
        \end{array}
    \]
    \caption{BSML primitives}
    \label{fig:bsml:primitives}
\end{figure}

Let us consider a function "f" that maps integers to values of type "'a" (denoted as "f: int->'a" in OCaml). The BSML primitive "mkpar f" produces a \emph{parallel vector} of type "'a par" when applied to function "f". Within this parallel vector, each processor, identified by the index value "i" within the range $0\leq i<$"bsp_p", stores the computed value of "f i". For instance, employing the expression "mkpar(fun i->i)" yields a parallel vector denoted as $\langle 0,~\ldots,~\mathtt{bsp\_p}-1\rangle$ of type "int par". Throughout subsequent discussions, we shall refer to this parallel vector as "this." Additionally, the function "replicate" possesses the type "'a -> 'a par" and can be defined as follows: "let replicate = fun x -> mkpar(fun i -> x)". By employing the expression "replicate x," the value "x" becomes uniformly available across all processors within the parallel vector. Parallel vectors always have size "bsp_p".

To apply a parallel vector of functions (which is not a function) to a parallel vector of values, one has to use the primitive "apply". Both "mkpar" and "apply" are executed within the pure computation phase of a super-step. For communications and an implicit synchronization barrier, the last two primitives "proj" and "put" should be applied. "proj" is essentially an inverse of "mkpar" but the resulting function is partial and only defined on the domain "[0, p-1]". As the first constant constructor of any inductively defined type is considered as the empty message, "put" allows to program any communication pattern of a BSP super-step. In the input vector of "put", each function encodes the message to be sent to other processors by the processor holding it. In the result vector, each function represents the message received from other processors by the processor holding the function.

Figure~\ref{fig:bsml:example} presents a small BSML example using its primitives and "parfun" which is part of its standard library. "List.map" and "List.fold_left" are part of the OCaml standard library and are sequential map and reduce functions.

Lines 4--5, we define a function "list_of_par" which converts a parallel vector into a list. This function requires a full super-step for its execution because it needs data exchanges. Also part of the BSML standard library, "procs" has type "int list" and is the list "[0;$\ldots$;bsp_p-1]".

Lines 7--8, we define an algorithmic skeleton: a parallel map that operates on a distributed list (represented here as a value of type "'a list par"). This function also requires the computation phase of a super-step and does not need any data exchange or synchronization.

Lines 10--13, we define the "reduce" algorithmic skeleton, using a binary associative operation "op" and a neutral element "e", it ``sums'' a distributed list into a single value. It proceeds in two steps. First, each processor compute a partial ``sum'' of the list it holds locally. Second, this vector of partial sums is transformed into a list which is finally summed up. As we call "list_of_par", a full super-step is required.

Finally, in lines 15--18, we implement a parallel function to solve the maximum prefix sum problem. Computing at the same time the maximum prefix sum and the sum of a list (in a pair) can be implemented using "map" and "reduce". For example, on a machine with at least 4 processors, the value of
"mps (mkpar(function|0->[1;2]|1->[-1;2]|2->[-1;3]|3->[-4]|_->[]))"
is "6". Indeed, the argument of "mps" is a distributed version (on 4 processors) of the list "[1;2;-1;2;-1;3;-4]" and its prefix with the largest sum is the list without its last element. We specify and prove the correctness of "mps" in Section~\ref{sec:application}.

\begin{figure}[!h]
    \centering
    \begin{minipage}{.85\textwidth}
\begin{lstlisting}[style=bsmlstyle,numbers=left,linerange={1-18},basicstyle=\codefigsize]
open Bsml
open Stdlib.Base

let list_of_par (v: 'a par) : 'a list = 
  List.map (proj v) procs

let map : ('a->'b) -> 'a list par -> 'b list par = 
  fun f v -> parfun (List.map f) v
    
let reduce : ('a->'a->'a) -> 'a -> 'a list par -> 'a = 
  fun op e v ->
  let locally_reduced = parfun (List.fold_left op e) v in 
  List.fold_left op e (list_of_par locally_reduced)

let mps : int list par -> int = 
  let op  (xm, xs) (ym, ys) = (max 0 (max xm (xs+ym)), xs+ys) in  
  let f x = (max 0 x, x) in
  fun v -> fst (reduce op (0, 0) (map f v))
\end{lstlisting}
    \end{minipage}
    \caption{A BSML Example}
    \label{fig:bsml:example}
\end{figure}

\section{Formalization of BSML Core and Standard Library}
\label{sec:bsmlwhy}
To be able to specify and write BSML programs, we need BSML primitives in WhyML. BSML primitives are implemented in parallel on top of MPI~\cite{snir98} called throught OCaml's Foreign Function Interface (FFI). Therefore, we cannot provide BSML in WhyML as an implementation. We need to give a BSML \emph{theory}: a set of constant, \emph{axioms} and function \emph{declarations}. The axiomatization of BSML primitives can be found in Figure~\ref{fig:why:primitives}. The semantics of functions \w{mkpar}, \w{apply}, \w{proj} and \w{put} are expressed in their contract (lines 12--24) while the strict positivity condition on \w{bsp_p} is given as an axiom on line 4. The type of parallel vector is abstract. Still we need to be able to observe parallel vectors. That is the role of logic function \w{get} which is a \w{ghost} function: it can only be used in specifications. A parallel vector is fully defined by the values all the processors hold as expressed by the axiom \w{extensionnality} in lines 9-10.
\begin{figure}[!h]
    \centering
    \begin{minipage}{.9\textwidth}
\begin{lstlisting}[style=whystyle,basicstyle=\codefigsize\ttfamily,numbers=left]
theory BSML
  use int.Int
  val constant bsp_p : int
  axiom at_least_one_processor : bsp_p > 0

  type par 'a       
  val ghost function get (_ : par 'a) (_ : int) : 'a
  axiom extensionality: 
    forall v v': par 'a. 
    (forall i: int. 0<=i<bsp_p -> get v i = get v' i) -> v = v'

  val mkpar (f : int -> 'a ) : par 'a 
    ensures { forall i:int. 0 <= i < bsp_p -> get result i = f i }
      
  val apply (f : par ('a -> 'b)) (v : par 'a ) : par 'b
    ensures { forall i:int. 0 <= i < bsp_p -> 
                       get result i = (get f i) (get v i) }
      
  val proj (v : par 'a) (x : int) : 'a 
    ensures { result = get v x } 
    
  val put (v : par (int -> 'a)) : par (int -> 'a) 
    ensures { forall s d:int. 0 <= s < bsp_p -> 0 <= d < bsp_p -> 
                             (get result d) s = (get v s) d }
end
\end{lstlisting}
    \end{minipage}
    \caption{BSML Theory in WhyML}
    \label{fig:why:primitives}
\end{figure}
The axiomatization is very close to the informal semantics of Figure~\ref{fig:bsml:primitives}. Instead of considering the parallel vectors globally with the notation $\pvec{v_0}{v_{p-1}}$, we consider each value $v_i$ denoted by \w{get v i}.

It is possible to realize this theory by a sequential implementation, for example implementing parallel vectors with sequential lists or arrays. This ensures the consistency of this theory.

To illustrate the use of this theory, we now specify, implement and verify several of the functions provided in the BSML standard library. The first one is \w{replicate}:
\begin{lstlisting}[style=whystyle,firstline=27, lastline=29]
  let replicate (x: 'a) : par 'a 
    ensures { forall i:int. 0 <= i < bsp_p -> get result i = x }
  = mkpar(fun _ -> x)
\end{lstlisting}
This verified function has only one post-condition: the result of replication is parallel vector which contains the same value everywhere.

In Section~\ref{sec:fbsp}, we mentioned the function "parfun" without defining it. Its implementation and specification follows, as well as the definition of function \w{parfun2}:
\begin{lstlisting}[style=whystyle,linerange={33-36,40-43}]
  let parfun (f : 'a -> 'b) (v: par 'a) : par 'b
    ensures { forall i:int. 0 <= i < bsp_p -> 
                       get result i = f (get v i) }
  = apply (replicate f) v
  let parfun2 (f : 'a -> 'b -> 'c) (u : par 'a) (v : par 'b) : par 'c
    ensures { forall i:int. 0 <= i < bsp_p -> 
                       get result i = f (get u i) (get v i) } 
  = apply (parfun f u) v
\end{lstlisting}
It shows how to use the \w{apply} primitive. There is also a \w{parfun3} function omitted here.

Next, we use the communication primitive \w{proj}. As we wrote in Section~\ref{sec:fbsp}, \w{proj} is essentially the inverse of \w{mkpar}. If we forget the cost of communication and synchronization, this function allows us to obtain the value of a vector \w{v} at a given processor \w{i}. However, it should not be used for such individual vector access, otherwise the performances would be extremely poor. The use of \w{proj} should be thought as a collective operation. Note that \w{proj} and \w{get} have the same semantics. However, the intent is very different: \w{get} is written only in specifications, can be thought as an indexed array access, and is used for \emph{local} reasoning,  while \w{proj} is used only in programs and requires a full super-step to execute. \w{proj} should rather be thought as a \emph{global} (i.e. concerning and involving all the processors) conversion of a parallel vector into a function.

To illustrate \w{proj}, we define the \w{list_of_par}. As we mentioned before this function requires a complete super-step to run. Again it should be seen as a \emph{global} conversion from parallel vectors to lists:
\begin{lstlisting}[style=whystyle,linerange={53-56}]
  let function list_of_par (v : par 'a) : list 'a 
    ensures { forall i:int. 0 <= i < bsp_p -> result[i] = get v i }
    ensures { length result = bsp_p }
  = map (proj v) (procs())
\end{lstlisting}

As in the BSML/OCaml version we call \w{procs} -- which needs to be a function  for \why to accept the code. \w{procs} returns the list of all processor identifiers. The definition of \w{procs} relies on a function \w{from_to} itself implemented using a \w{init} function. Our contribution does also contain a library of sequential functions, mostly on lists, as well as verified lemmas stating their properties. These functions can in most cases be used both in programs and in specifications.

Finally, the \w{put} primitive is illustrated to implement a broadcast function. This data exchange (and implicit global synchronization) function is more precise than \w{proj}. We remind that after a \w{put}, for all processors \w{d} and \w{s}, the result function at destination processor "d", applied to the identifier of source processor "s" retuns the value of the input function at source processor "s" applied to destination processor "d".

The definition of the "bcast_direct" function of the standard library follows. This function is used to broadcast a value from a \w{root} processor to all other processors. To do so, first, we prepare a function vector for the processors to make the messages to send to each other (local definitions \w{make_msg} and \w{to_send}). It is clear that only the \w{root} processor with send data. The other processors' message is \w{None} which is interpreted by the BSML/OCaml implementation as an empty message. Second, the local definition \w{received} proceeds with the data exchange and ends the super-step. \w{received} is a parallel vector of functions. What we are interested in is the value sent by processor \w{root}. That is why the local definition \w{optional_result} then applies this parallel vector of functions to the replicated value \w{root}. Of course, the obtained message is encapsulated in a \w{Some} constructor. Therefore, all the processors finally apply \w{remove_option} to yield the final result. The broadcast is meaningless if \w{root} is not a valid processor identifier. In this case, the exception \w{Bcast} is raised:

\begin{lstlisting}[style=whystyle,linerange={215-228}]
  let bcast_direct (root : int) (v : par 'a)
    ensures { 0 <= root < bsp_p -> 
              forall i:int. 0 <= i < bsp_p -> 
                       get result i = get v root }
    raises { Bcast }
  = if (0 <= root) && (root < bsp_p )
    then
      let make_msg src load = fun _ -> 
          if src = root then Some load else None in 
      let to_send  = apply (mkpar make_msg) v in 
      let received = put to_send in 
      let optional_result = apply received (replicate root) in 
      parfun remove_option optional_result
    else raise Bcast
\end{lstlisting}

Our BSML theory allows us to write BSML programs and their specifications and is expressive enough for the \why3 framework to verify that they indeed satisfy their specifications.

We only presented a sub-set of the functions of the BSML standard library we implemented, and we refer to the companion artifact for the complete set of functions. For example, we also provide the "shift", "shift_right" and "shift_left" communication functions, which offer a different communication pattern than \w{bcast_direct}: Each data item is shifted by a certain number of processors.

\section{Verified Scalable Maximum Prefix Sum}
\label{sec:application}
\begin{figure}
    \centering
    \begin{minipage}{.9\textwidth}
\begin{lstlisting}[style=whystyle,basicstyle=\codefigsize\ttfamily,numbers=left,linerange={24-43}]
  let map_par (f: 'a->'b) (dl: par (list 'a)) : par (list 'b)
    ensures { forall i:int. 0 <= i < bsp_p ->
                       get result i = map f (get dl i) }
    ensures { to_list result = map f (to_list dl) }
  = let ghost _ = flatten_map f (list_of_par dl) in
    parfun (map f) dl
    
  let reduce_par (ghost inv: 'a->bool)
    (op: 'a->'a->'a) (e: 'a) (dl: par(list 'a)) : 'a 
    requires { associative inv op }
    requires { neutral inv op e }
    requires { preserves inv op }
    requires { inv e }
    requires { forall i:int. 0 <= i < bsp_p -> 
                        satisfies inv (get dl i) }
    ensures { result = fold_left op e (to_list dl) }
  = let ghost _ = fold_left_flatten inv op e (list_of_par dl) in  
    let reduce_seq l = fold_left op e l in 
    let partial_reductions = parfun reduce_seq dl in 
    reduce_seq (list_of_par partial_reductions)
\end{lstlisting}
    \end{minipage}
    \caption{Verified Algorithmic Skeletons in WhyML}
    \label{fig:bsml:why:skeletons}
\end{figure}

\begin{figure}
    \centering
    \begin{minipage}{.9\textwidth}
\begin{lstlisting}[style=whystyle,basicstyle=\codefigsize\ttfamily,numbers=left,linerange={30-42}]
  predicate associative (inv: 'a->bool) (op: 'a->'a->'a) = 
    forall x y z:'a. inv x -> inv y -> inv z ->
      op x (op y z) = op (op x y) z
  
  predicate neutral (inv: 'a->bool) (op: 'a->'a->'a) (e: 'a) =
    (forall x:'a. inv x -> op x e = x) /\ 
    (forall x:'a. inv x -> op e x = x)
    
  predicate preserves (inv: 'a->bool) (op: 'a->'a->'a) = 
    forall x y : 'a. inv x -> inv y -> inv (op x y)

  predicate satisfies (inv: 'a->bool) (l: list 'a) = 
    forall i:int. 0 <= i < length l -> inv (nth i l)
\end{lstlisting}
    \end{minipage}
    \caption{Algebra Concepts}
    \label{fig:bsml:why:algebra}
\end{figure}

To exercise the formalization presented in the previous section, we specify and verify an implementation of the maximum prefix sum informally presented in Section~\ref{sec:fbsp}. As in the BSML implementation, the implementation with WhyML relies on algorithmic skeletons. The skeleton \w{par_map} is defined in lines 1--6 of Figure~\ref{fig:bsml:why:skeletons}. The only different with its BSML/OCaml counterpart is the post-conditions including one expressed as a correspondence with the sequential \w{map}. Given a distributed list \w{dl} (of type \w{par(list 'a)}), one obtains the same result by either applying \w{map_par} then transforming the obtained distributed list into a list with \w{to_list}, or applying the sequential \w{map} to the sequentialization of the distributed list. Line 5 is just a hint for the provers: an application of lemma \w{flatten_map} that basically commute \w{map} and \w{flatten}.

The implementation (lines 8-20) of the parallel reduction \w{reduce_par} is also very close to its BSML/OCaml counterpart of Figure~\ref{fig:bsml:example}. As expected, the post-condition on line 16 is expressed with respect to the sequential reduction here implemented with the usual \w{fold_left} function. As the result is already a sequential value there is no need to sequentialize it. However, this correspondence is true only if \w{op} is associative and \w{e} is its neutral element which are two pre-conditions stated lines 10--11. There are two additional pre-conditions and a \w{ghost} argument, i.e. an argument only used in the contract (and possible annotations) of the function. The reason is again to deal with a form of partial functions. \w{op} is a total function, but it may not have the desired properties (associativity, neutral element) on all the values of its input type. Indeed, the OCaml version of "op" for "mps" that we will also use in the WhyML version of \w{mps}, is not associative if we consider all pairs of integers. In the maximum prefix sum problem, the first component of such a pair represents the maximum prefix sum, it is therefore positive, and the second component the sum of the list, thus it is lower or equal to the first component. The ghost argument \w{inv} expresses such properties on the values manipulated during the reduction. This is an invariant: \w{op} should preserve the property (line 12) and the input values \w{e} and \w{dl} should satisfy this property (line 13). The predicates \w{associative}, \w{neutral}, \w{preserves} and \w{satisfies} are defined in Figure~\ref{fig:bsml:why:algebra}. Such definitions work also well when there is no need for an invariant: in this case we simply use the constant boolean function always returning \w{true}.

\begin{figure}[!h]
    \centering
    \begin{minipage}{.9\textwidth}
\begin{lstlisting}[style=whystyle,basicstyle=\codefigsize\ttfamily,numbers=left,linerange={29-30,43-53,128-133,139-143}]
  let function mps_spec (l: list int) : int 
  = maximum (map sum (prefix l))

  function ms (l: list int) : (int,int) 
  = (mps_spec l, sum l)

  let function op (x: (int,int)) (y: (int,int)) : (int, int)
  = (max (max 0 (fst x)) ((snd x)+(fst y)), (snd x)+(snd y)) 

  let function f (x : int) = (max 0 x, x) 

  predicate mps_sum_inv (x: (int,int)) 
  = 0 <= fst x /\ fst x >= snd x

  let function mps_seq (l: list int) : int 
    ensures { result = mps_spec l } 
  = let ghost _ = first_homomorphism_theorem ms op l in
    assert { fold_left op (0, 0) (map f l) = ms l };
    fst (fold_left op (0, 0) (map f l))  

  let mps_par (dl: par(list int)) : int 
    ensures { result = mps_spec (to_list dl) } 
  = let mapped = map_par f dl in
    fst (reduce_par mps_sum_inv op (0, 0) mapped)  
\end{lstlisting}
    \end{minipage}
    \caption{Verified Maximum Prefix Sum in WhyML}
    \label{fig:bsml:why:mps}
\end{figure}

With these skeletons, it is possible to implement a parallel function to compute the maximum prefix sum of a distributed list as we did in Section~\ref{sec:fbsp}. First, we define a \emph{specification} as an inefficient function but direct translation of the informal specification: the \w{mps_spec} function on lines 1--2 of Figure~\ref{fig:bsml:why:mps}. We also define \w{op} (lines 7--8) and \w{f} (line 10) which are the arguments to \w{map} and \w{reduce} as in the BSML/OCaml example of Figure~\ref{fig:bsml:example}. This time they are not local definitions because we need to state and verify some lemmas about them and because we have two versions of \w{mps}: \w{mps_seq} and \w{map_par}.
The invariant explained above is defined lines 12--13. We need an auxiliary function to verify the correctness of our functions with respect to the specification: \w{ms} (line 4--5) is the tupling of \w{mps_spec} and \w{sum}. The rest of the code in Figure~\ref{fig:bsml:why:mps} is the definitions of the sequential and parallel versions of the maximum prefix sum computation. Both of them are expressed as a composition of map and reduce.

The proof that \w{mps_seq} indeed implements the specification \w{mps_spect} proceeds by using the first homomorphism theorem. This theorem states that a homomorphic function can be implemented as a composition of map and reduce. A function \w{f} is homomorphic when there exists a binary operation $\odot$ such that: \w{forall l1 l2: list 'a. f(l1++l2) = (f l1)} $\odot$ \w{(f l2)} where \w{++} denotes list concatenation. \w{mps_spec} is not homomorphic but \w{ms} is. Two lines of annotations are necessary to guide the provers in the sequential case (lines 17--18). The parallel case does need any annotation: basically the contracts of \w{map_par} and \w{reduce_par} state their correspondence with their sequential counterpart thus the correspondence of the parallel \w{mps_par}— with the sequential \w{mps_seq}, and \w{mps_seq} satisfies \w{mps_spec}.

The full development is about 600 lines of WhyML with about 45\% of specifications and 55\% of code. It generates 74 goals, 100\% of which are proved. Their verification produces 37 sub-goals. The strategy \texttt{Auto level 2} is used: it tries the provers CVC4, Alt-Ergo, CVC5 and Z3 with a short timeout (1s). If the goal is not proved then it splits the goal and try on the sub-goals with the same timeout and finally if necessary tries with a larger timeout (10s). Alt-Ergo version 2.4.3 proved 11 goals taking between 0.02s and 0.56s (when successful) and CVC4 version 1.6 proved 91 goals taking between 0.04s and 2.45s. Several sub-goals can contribute to a goal to be proved. For example the verification condition of \w{mps_seq} is split in 3 sub-goals. In the number of the goals proved by CVC4 and Alt-Ergo the root goals verified because their sub-goals are proved are not counted. In our case, only 9 goals needed to be split to achieve their proofs.

\section{Related Work}
\label{sec:related_work}
BSP-WHY~\cite{FG2010:HLPP,FG2015:IJPP} also uses (a previous version of) \textsc{Why} to verify bulk synchronous parallel programs. However, the two approaches are very different. BSP-WHY considers BSP programs written in an imperative style close to BSPlib~\cite{HILL1998:BSPLIB}. The verification proceeds by transforming well-formed programs --- a sub-class of what has been formally defined later by Dabrowski as textually aligned programs~\cite{DAB2019:JLAMP} --- into sequential simulating programs that are then verified using \textsc{Why}. The BSP-WHY code cannot be run on parallel machines.

The work closest to our is the specification, verification and extraction of BSML programs using the Coq proof assistant. Early contributions started with the work of Gava~\cite{GAV2003:PPL}. A formalization of BSML primitives in a style very close to the \why formalization presented in this paper was proposed by Tesson and Loulergue~\cite{TL2011:ICCS} and used in a framework, named \textsc{SyDPaCC}, for the verification of BSP functional programs~\cite{ELT2014:ITP,LRT2014:SAC}. The two main differences with our work is that: proofs are much less automated in Coq than in \why but the framework leverages the type-class resolution mechanism of Coq to automatically parallelize programs. For example in this framework, the user does not need to write the code for \w{mps_seq} and \w{mps_par}, but only needs to write \w{mps_spec} and to prove that its tupling with \w{sum} is leftwards and rightwards (i.e. can be written as calls to \w{fold_left} and \w{fold_right}) and exhibits a weak right inverse. The framework would then use transformation theorems to automatically obtain \w{mps_seq} and then verified correspondences as expressed in the post-conditions of \w{map_par} and \w{reduce_par} to automatically produce \w{mps_par}~\cite{LBT2015:IJPP}.

Ono et al.~\cite{OHT2011:SEFM} employed Coq to verify Hadoop MapReduce programs and extract Haskell code for Hadoop Streaming or directly write Java programs annotated with JML, utilizing Krakatoa~\cite{FM2007:CAV} to generate Coq lemmas. The first part of their work is functional and therefore closest to our work. However, it is limited to MapReduce which is more general than the \w{map_par} and \w{reduce_par} skeletons but is less expressive than BSML. The second part of their work is more imperative.

\section{Conclusion and Future Work}
\label{sec:conclusion}
We were able to formalize the primitives of the parallel programming library BSML with WhyML and leverage \why for verifying a large part of the BSML standard library as well as an application written in BSML. We plan to experiment the extracted code more thoroughly and on larger parallel machines with a few thousand cores.

WhyML offers exceptions and references thus allows to write imperative programs. However, such programs cannot be passed as arguments to higher-order functions. It therefore limits the usage of imperative features with BSML as all primitives are higher-order functions. The code outside BSML primitives can be imperative thus the sequencing of BSP super-steps could be imperative. It is also possible to use imperative features to implement pure functions passed as arguments to BSML primitives. Also, it is possible to deal with partial functions as we did with "remove_some". We plan to explore all these possibilities in the future.
\bibliographystyle{splncs04}
\bibliography{bibliography,references}
\end{document}